\documentclass[showpacs,prl]{revtex4-1}
\usepackage{amsfonts}
\usepackage{graphicx}
\usepackage{amsmath}
\usepackage{amssymb}

\begin{document}

\title{Vanishing of interband light absorption in a persistent spin helix state}
\author{Zhou Li$^1$}
\author{F. Marsiglio$^{2}$}
\author{J. P. Carbotte$^{1,3}$}

\affiliation{$^1$ Department of Physics, McMaster University,
Hamilton, Ontario,Canada,L8S 4M1 \\
$^2$ Department of Physics, University of Alberta, Edmonton,
Alberta,T6G 2E1\\
$^3$ Canadian Institute for Advanced Research, Toronto, Ontario,
Canada M5G 1Z8}

\begin{abstract}
Spin-orbit coupling plays an important role in various properties of very different
materials. Moreover efforts are underway to control the degree and quality of
spin-orbit coupling in materials with a concomitant control of transport properties.
We calculate the frequency dependent optical conductivity in systems
with both Rashba and Dresselhaus spin-orbit coupling. We find that
when the linear Dresselhaus spin-orbit coupling is tuned to be equal
to the Rashba spin-orbit coupling, the interband optical
conductivity disappears. This is taken to be the signature of the
recovery of SU(2) symmetry. The presence of the cubic Dresselhaus
spin-orbit coupling modifies the dispersion relation of the charge
carriers and the velocity operator. Thus the conductivity is
modified, but the interband contribution remains suppressed at most
but not all photon energies for a cubic coupling of reasonable
magnitude. Hence, such a measurement can serve as a diagnostic probe of engineered spin-orbit coupling.
\end{abstract}

\date{\today }

\maketitle

Spin-orbit coupling in semiconductors \cite{wolf01} and at the
surface of three dimensional topological insulators
\cite{Hasan,Moore1,Qi,Bernevig,Fu,Chen1,Li3} where protected
metallic surface states exist, plays a crucial role in their
fundamental physical properties. Similarly pseudospin leads to novel
properties in graphene \cite{Mac,Novo2,Zhang}  and other two
dimensional membranes, such as single layer $MoS_{2}$
\cite{Mak1,Splendiani,Lebegue,Lee,Li12,Li13} and silicene
\cite{Drum,Aufray,Stille,Ezawa1,Ezawa2}. In particular $MoS_{2}$ has
been discussed within the context of valleytronics where the valley
degree of freedom can be manipulated with the aim of encoding
information in analogy to spintronics. Spin-orbit coupling has also
been realized in zincblende semiconductor quantum wells
\cite{awschalom09,Walser,bernevig06} and neutral atomic
Bose-Einstein condensates \cite{lin11} at very low temperature
\cite{bloch08}.

\begin{figure}[tp]
\begin{center}
\includegraphics[height=6.2in,width=6.2in]{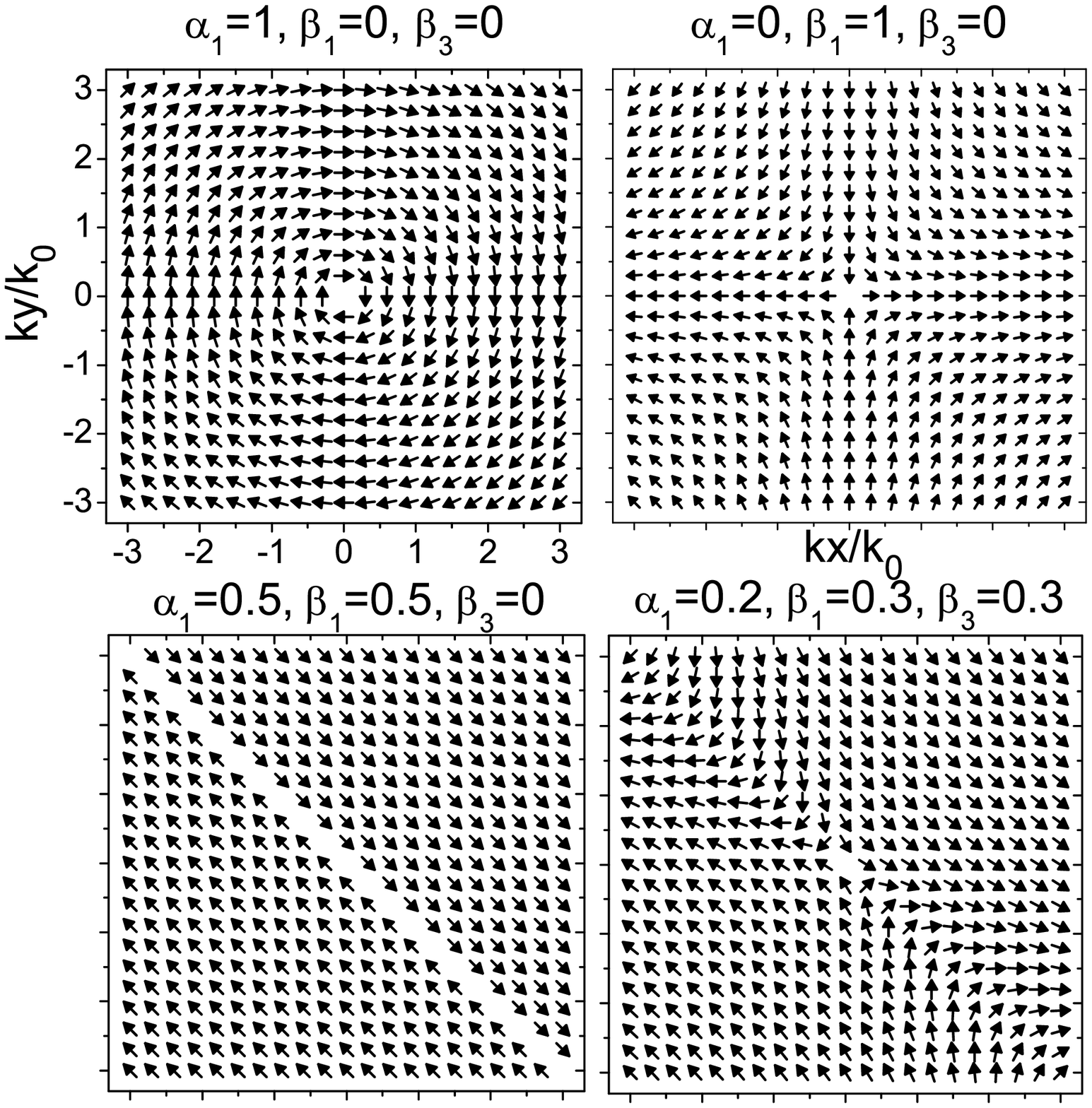}
\end{center}
{Fig.1. Spin texture in the conduction band as a function of
momentum $k_{x}/k_{0}$, $k_{y}/k_{0}$ for various values of Rashba
($\alpha_1$), Dresselhaus ($\beta_{1}$), and cubic Dresselhaus
($\beta_{3}$) spin-orbit coupling. In the case of purely Rashba
coupling (upper left frame), the spin is locked in the direction
perpendicular to momentum, while for linear Dresselhaus coupling
(upper right frame) the y-component of spin is of opposite sign to
that of its momentum. For the persistent spin helix state (lower
left frame) all spins are locked in the $3\pi/4$ direction and
oppositely directed on either side of this critical direction. The
lower right frame shows the spin texture for a case with all three
kinds of coupling. } \label{fig1}
\end{figure}

\begin{figure}[tp]
\begin{center}
\includegraphics[height=6.6in,width=4.0in,angle=-90]{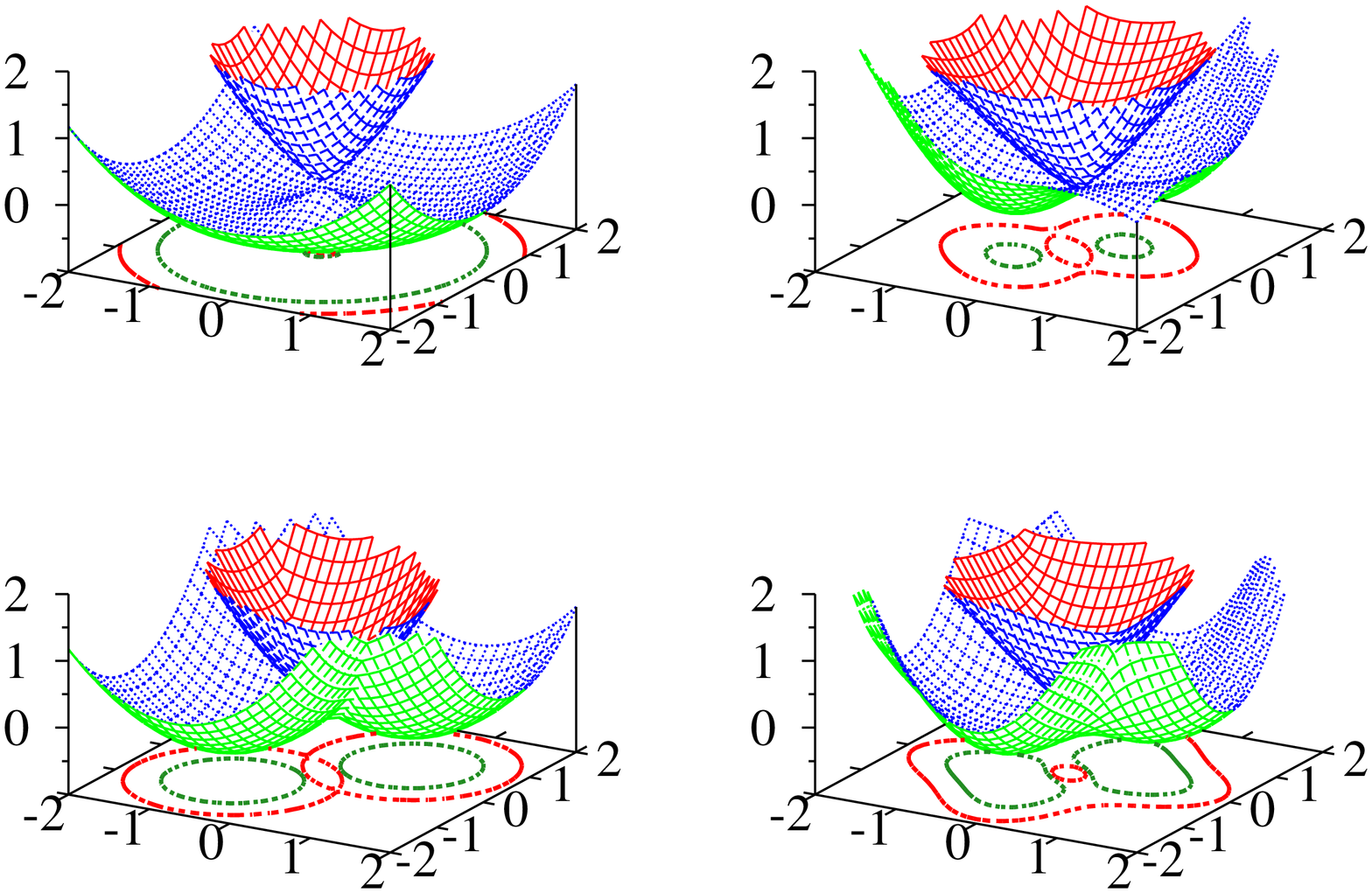}
\end{center}
{Fig.2. Band structure of the conduction and valence band (
Eq.~(\ref{eigenvalues})) as a function of momentum $k_{x}/k_{0}$,
$k_{y}/k_{0}$ for various values of Rashba ($\alpha_1$), Dresselhaus
($\beta_{1}$), and cubic Dresselhaus ($\beta_{3}$) spin-orbit
coupling. The left two panels are for pure Rashba $\alpha_1=1.0,
\beta_1=0.0, \beta_{3}=0.0$ (top panel) and Rashba equals to
Dresselhaus $\alpha_1=0.5, \beta_1=0.5, \beta_{3}=0.0$ (bottom
panel). The right two panels are for $\alpha_1=0.4, \beta_1=0.4,
\beta_{3}=0.3$ (top panel) and $\alpha_1=0.2, \beta_1=0.8,
\beta_{3}=0.3$ (bottom panel). The dispersion curves are profoundly
changed from the familiar Dirac cone of the pure Rashba case when
$\beta_1$ and $\beta_{3}$ are switched on. In the contour plots, red
refers to energy $0.2E_0$ and dark green refers to energy $-0.2E_0$.
} \label{fig2}
\end{figure}

\begin{figure}[tp]
\begin{center}
\includegraphics[height=6.2in,width=5.5in]{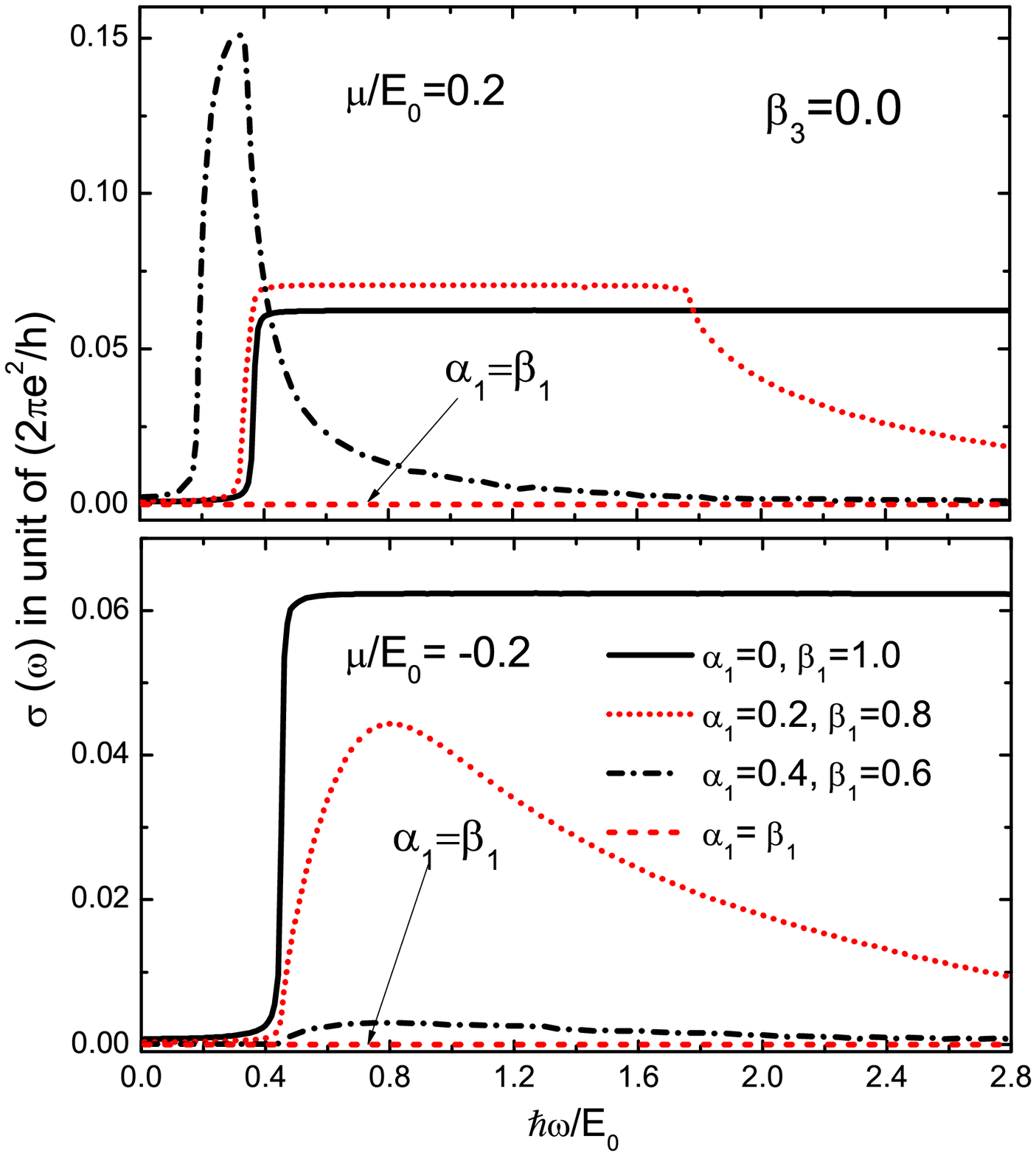}
\end{center}
{ Fig.3. The interband contribution to the longitudinal optical
conductivity of Eq.~(\ref{Cond}) for various values of $\alpha_{1}$
and $\beta_{1}$ as labeled, with $\beta_3$ set to zero. In the top
frame the chemical potential was set at $\mu/E_{0}=0.2$ and in the
bottom $\mu/E_{0}=-0.2$.} \label{fig3}
\end{figure}

\begin{figure}[tp]
\begin{center}
\includegraphics[height=6.5in,width=6.5in]{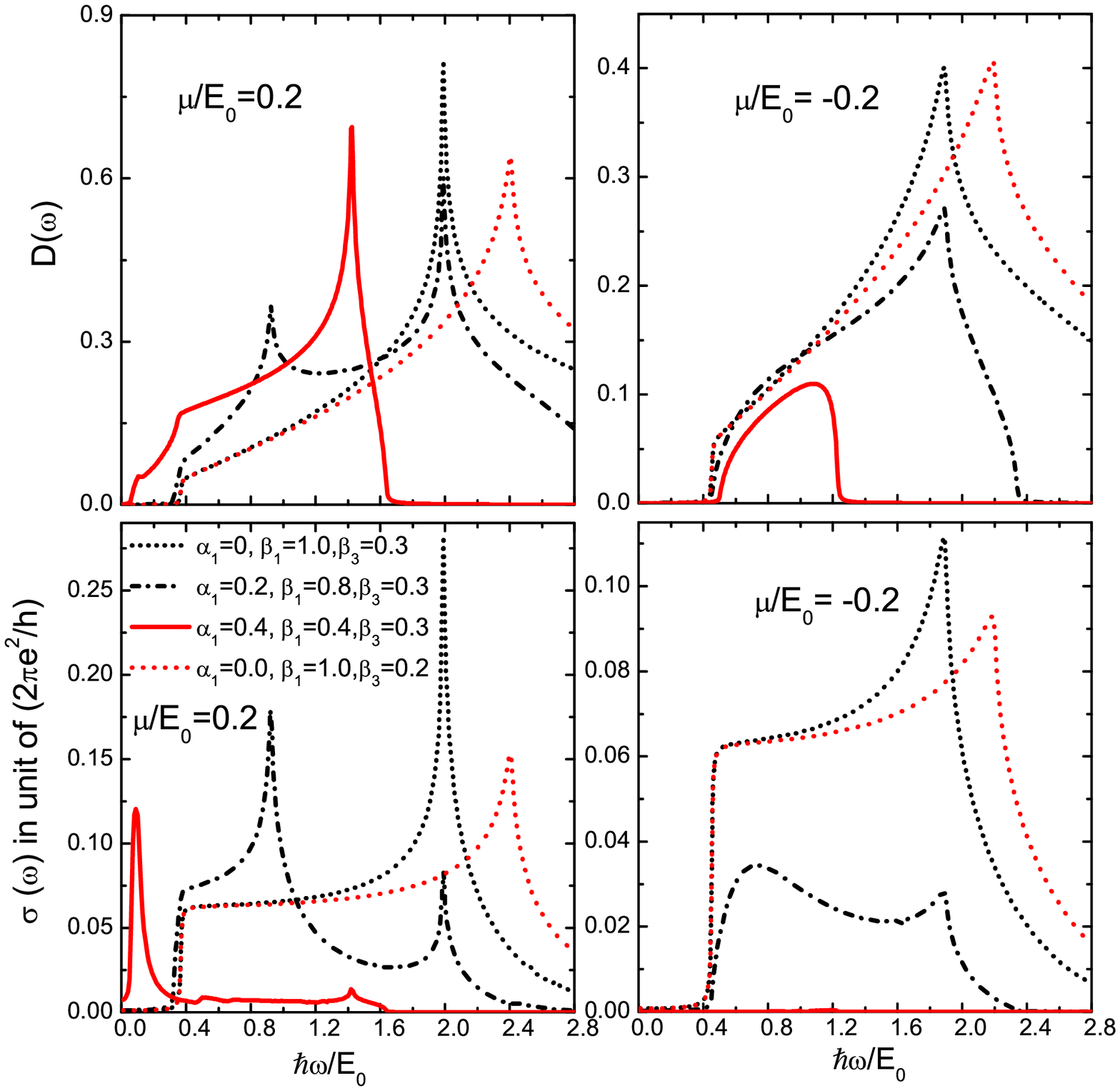}
\end{center}
{Fig.4. Joint density of states $D(\omega)$ (top two panels) defined
in Eq.~(\ref{DOS}) which involves the same transitions as does the
interband conductivity (bottom two panels) of Eq.~(\ref{Cond}) but
without the critical weighting
$\frac{(V_{x}S_{2}+V_{y}S_{1})^{2}}{(S_{1}^{2}+S_{2}^{2})\omega}$.
Left column is for positive chemical potential $\mu/E_{0}=0.2$ and
the right for -0.2.} \label{fig4}
\end{figure}

\begin{figure}[tp]
\begin{center}
\includegraphics[height=7.4in,width=4.2in]{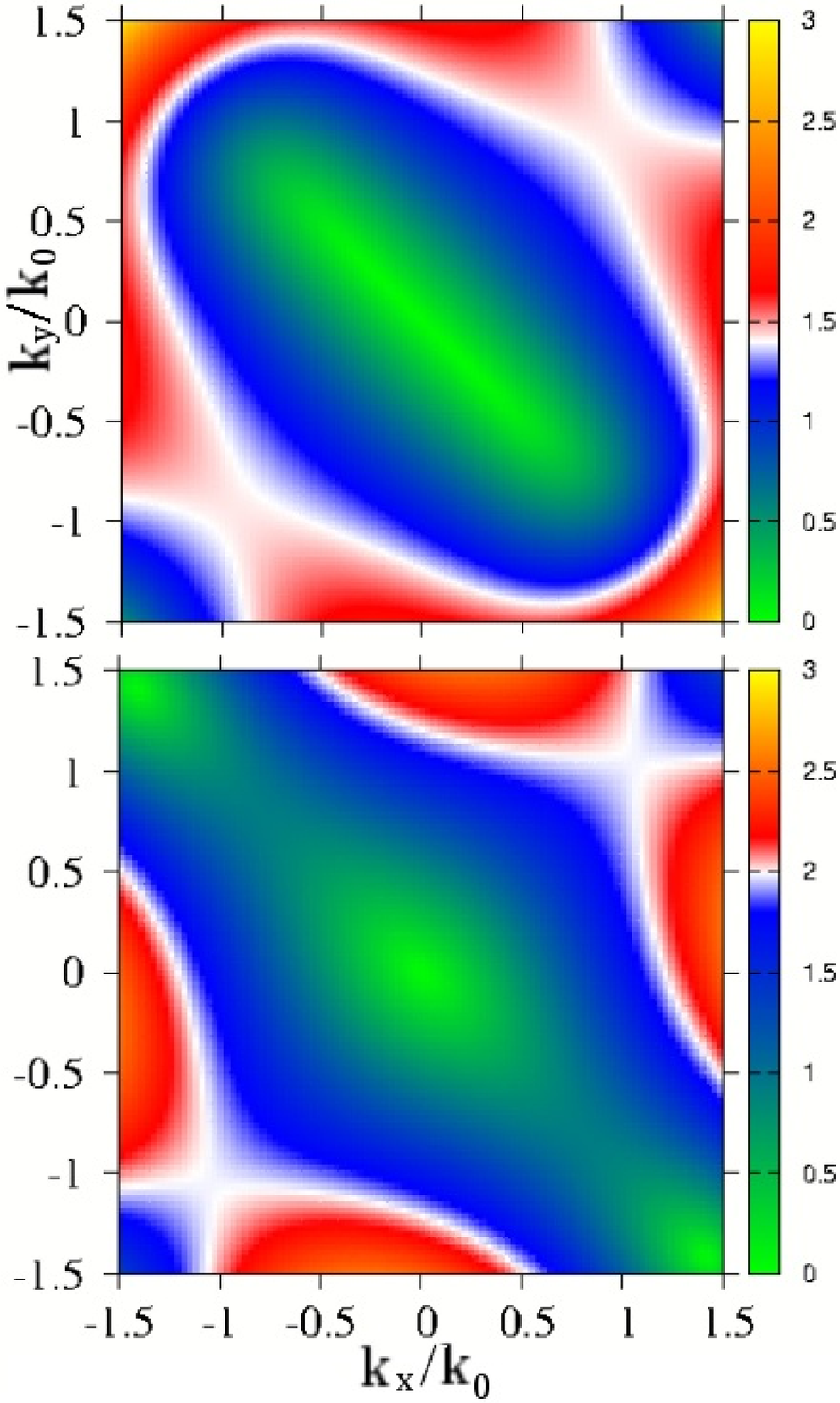}
\end{center}
{Fig.5. Color contour plot of the energy difference
$2\sqrt{S_{1}^{2}+S_{2}^{2}}\equiv E_{+}-E_{-}$, as a function of
momentum ($k_{x},k_{y}$) in units of $k_{0}$ for $\alpha_1=0.4,
\beta_1=0.4, \beta_{3}=0.3$ (top panel) and $\alpha_1=0.2,
\beta_1=0.8, \beta_{3}=0.3$ (bottom panel). } \label{fig5}
\end{figure}

In some systems both Rashba \cite{rashba60} and
Dresselhaus \cite{dresselhaus} spin-orbit coupling are manipulated,
the former arising from an inversion asymmetry of the grown layer
while the latter comes from the bulk crystal. In general spin-orbit
coupling will lead to rotation of the spin of charge carriers as
they change their momentum, because SU(2) symmetry is broken. In
momentum space this has been observed by angle-resolved
photoemission spectroscopy (ARPES) as the phenomenon of spin
momentum locking. In a special situation when the strength of Rashba
and Dresselhaus spin-orbit coupling are tuned to be equal, SU(2)
symmetry is recovered and a persistent spin helix state is found
\cite{awschalom09,Walser,bernevig06}. This state is robust against
any spin-independent scattering. However it will be potentially destroyed by the
cubic Dresselhaus term which is usually tuned to be negligible.

To describe these effects we consider a model Hamiltonian
describing a free electron gas with kinetic energy given simply by
${\hbar ^{2}k^{2}}/(2m)$, which describes charge carriers with effective mass
$m$. We also include spin-orbit coupling terms,
with linear Rashba ($\alpha_1 $) and Dresselhaus ($\beta _{1}$) couplings, along with a
cubic Dresselhaus ($\beta_{3}$) term.
The Hamiltonian is
\begin{eqnarray}
\hat{H}_{0}=\frac{\hbar ^{2}k^{2}}{2m} \hat{I}+\alpha_1
(k_{y}\hat{\sigma} _{x}-k_{x}\hat{\sigma} _{y}) +\beta
_{1}(k_{x}\hat{\sigma} _{x}-k_{y}\hat{\sigma} _{y})-\beta
_{3}(k_{x}k_{y}^{2}\hat{\sigma} _{x}-k_{y}k_{x}^{2}\hat{\sigma}
_{y}). \label{H0}
\end{eqnarray}
Here $\hat{\sigma} _{x},\hat{\sigma} _{y}$ and $\hat{\sigma} _{z}$
are the Pauli matrices for spin (or pseudospin in a neutral atomic
Bose-Einstein condensate) and $\hat{I}$ is the unit matrix. For
units we use a typical wave vector $k_{0}\equiv
m\alpha_{0}/\hbar^{2}$ with corresponding energy
$E_{0}=m\alpha^{2}_{0}/\hbar^{2}$, where $\alpha_{0}$ is a
representative spin-orbit coupling which has quite different values
for semiconductors ($\alpha_{0}/\hbar\approx10^{5}m/s$, estimated
from Ref. \cite{bernevig06}) and cold atoms
($\alpha_{0}/\hbar\approx0.1m/s$, estimated from Ref. \cite{lin11}).
The mass of a cold atom is at least 1000 times heavier than that of an
electron and the wavelength of the laser used to trap the atoms is
at least 1000 times (estimated from Ref. \cite{lin11}) larger than
the lattice spacing in semiconductors.

In this report we study the dynamic longitudinal optical
conductivity of such a spin-orbit coupled 2D electron gas. We find
that the interband optical absorption will disappear when the Rashba
coupling is tuned to be equal to the Dresselhaus coupling strength.
We discuss the effect of nonlinear (cubic) Dresselhaus coupling on
the shape of the interband conductivity and the effect of the
asymmetry between the conduction and valence band which results from
a mass term in the dispersion curves.

\section{Results}

We compute the optical conductivity (see Methods section) as a
function of frequency, for various electron fillings and spin-orbit
coupling strengths. In all our figures we will use a dimensionless
definition of spin-orbit coupling; for example, the choice of values
designated in the lower right frame of Fig.~1, $\alpha_1=0.2,
\beta_1=0.3$, and $\beta_{3}=0.3$, really means
$\alpha_1/\alpha_{0}=0.2,\beta_1/\alpha_{0}=0.3$, and
$\beta_{3}k^{2}_{0}/\alpha_{0}=0.3$.

In Fig.~1 we plot the spin direction in the conduction band as a
function of momentum for several cases. The top left frame is for
pure Rashba coupling, in which case spin is locked to be
perpendicular to momentum \cite{Hasan} as has been verified in spin
angle-resolved photoemission spectroscopy studies
\cite{Lanzara1,Lanzara2,Xu,Wang}. The top right frame gives results
for pure linear Dresselhaus coupling (no cubic term $\beta_{3}=0$).
The spin pattern is now quite different; the direction of the spin
follows the mirror image of the momentum about the x-axis. The lower
left frame for equal linear Rashba and Dresselhaus coupling is the
most interesting to us here. All spins are locked in one direction,
namely $\theta=3\pi/4$ with those in the bottom (upper) triangle
pointing parallel (anti-parallel) to the $3\pi/4$ direction,
respectively. This spin arrangement corresponds to the persistent
spin helix state of Ref. \cite{awschalom09,Walser,bernevig06}. The
condition $\alpha_1 = \beta_1$ and $\beta_{3}=0$ is a state of zero
Berry phase
\cite{Shen1} and was also characterized by Li {\it et al.} 
\cite{Shen} as a state in which the spin transverse ``force" due to
spin-orbit coupling cancels exactly. Finally the right lower frame
includes a contribution from the cubic Dresselhaus term of
Eq.~(\ref{H0}) and shows a more complex spin arrangement. Spin
textures have been the subject of many recent studies
\cite{Lanzara1,Lanzara2,Xu,Wang,Khom}. In Fig.~2 we present results
for the dispersion curves in the conduction and valence band
$E_{+/-}(k)$ of Eq.~(\ref{eigenvalues}) as a function of momentum
$k$. The two left panels are pure Rashba (top) and Rashba equals to
Dresselhaus (bottom, see also Fig.1 of Ref.\cite{Mars} where only
the contour plots of the valence band is shown). The two right
panels include the Dresselhaus warping cubic term which profoundly
affects the band structure.

The optical conductivity is obtained through transitions from one
electronic state to another. In general these can be divided into
two categories --- transitions involving states within the same
band, and interband transitions. Here we focus on interband
transitions; the interband optical conductivity is given by
\begin{equation}
\sigma _{xx}(\omega )=\frac{e^{2}}{i\omega }\frac{1}{4\pi ^{2}}
\int_{0}^{k_{cut}}kdkd\theta \frac{(V_{x}S_{2}+V_{y}S_{1})^{2}}{S_{1}^{2}+S_{2}^{2}}
\biggl[\frac{f(E_{+})-f(E_{-})}{\hbar \omega -E_{+}+E_{-}+i\delta
} - \frac{f(E_{+})-f(E_{-})}{\hbar \omega -E_{-}+E_{+}+i\delta }\biggr],
\label{Cond}
\end{equation}
where $f(x)=1/(e^{(x-\mu)/k_{B}T}+1)$ is the Fermi-Dirac
distribution function with $\mu$ the chemical potential. For $\beta
_{3}=0$ and $\beta _{1}=\alpha_1 $, we have a cancellation in the
optical matrix element, $V_{x}S_{2}+V_{y}S_{1}=0 $; remarkably the
interband contribution vanishes. This result is central to our work
and shows that in the persistent spin helix state the interband
contribution to the dynamic longitudinal optical conductivity
vanishes. This is the optical signature of the existence of the spin
helix state which exhibits remarkable properties. With  $\beta
_{3}=0$ the optical matrix element is $(\beta _{1}^{2} -
\alpha_1^{2})k_{y}/\hbar$. Thus, pure Rashba or pure (linear)
Dresselhaus coupling will both lead to exactly the same conductivity
although the states (and spin texture) involved differ by a phase
factor of $\pi$. When they are both present in equal amounts this
phase leads to a cancelation which reduces the interband transitions
to zero as the two contributions need to be added before the square
is taken. Of course the joint density of states, widely used to
discuss optical absorption processes, remains finite. It is given by
\begin{equation}
D(\omega )=\frac{1}{4\pi ^{2}}\int_{0}^{k_{cut}}kdkd\theta \ [f(E_{+})-f(E_{-})]
Im \biggl[\frac{1}{\hbar \omega -E_{+}+E_{-}+i\delta} -
\frac{1}{\hbar \omega -E_{-}+E_{+}+i\delta }\biggr]
\label{DOS}
\end{equation}
and will be contrasted with the interband optical conductivity
below.

We first focus on the case $\beta_3 = 0$. The interband conductivity
is shown in Fig.~3 as a function of frequency for positive (top
frame) and negative (bottom frame) chemical potential ($\mu/E_0 =
\pm 0.2$). It is clear that there is a considerable difference
between the two cases, and there is also considerable variation with
the degree of Rashba vs. Dresselhaus coupling. This will be
discussed further below. Most important is that for equal amounts of
Rashba and Dresselhaus coupling, the interband conductivity is
identically zero for all frequencies.

What is the impact of a finite value of $\beta_3$ ? In Fig.~4 we
show both the joint density of states (top two panels) and the
interband conductivity (bottom two panels) for non-zero $\beta_3$
for $\mu/E_0 = 0.2$ (left panels) and $\mu/E_0 = -0.2$ (right
panels). Various combinations of $\alpha_1 $, $\beta_{1}$ and
$\beta_{3}$ are shown as labeled on the figure. There is a striking
asymmetry between positive and negative values of the chemical
potential. This asymmetry has its origin in the quadratic term
${\hbar^{2}k^{2}}/{(2m)}$ of the Hamiltonian (\ref{H0}) which adds
positively to the energy in both valence and conduction band while
the Dirac like contribution is negative ($s=-1$) and positive
($s=+1$) respectively [see Eq. (\ref{eigenvalues})]. While the
quadratic piece drops out of the energy denominator in
Eq.~(\ref{Cond}) it remains in the Fermi factors $f(E_+)$ and
$f(E_-)$.

Several features of these curves are noteworthy. They all have van
Hove singularities which can be traced to extrema in the energy
difference $E_{+}-E_{-}=2\sqrt{S_{1}^{2}+S_{2}^{2}}$. Taking $\beta
_{3}=0$ for simplicity, this energy becomes $2k\sqrt{\alpha_{1}
^{2}+\beta_{1}^{2}+2\alpha_1 \beta _1 \sin (2\theta )}$ which
depends on the direction ($\theta$) of momentum $\mathbf{k}$, but
has no minimum or maximum as a function of $|\mathbf{k}|=k$. To get
an extremum one needs to have a non-zero cubic Dresselhaus term.
This gives dispersion curves which flatten out with increasing
values of $k$. The dependence of the energy $E_{+}-E_{-}$ on
momentum is illustrated in Fig.~5 where we provide a color plot for
this energy as a function of $k_{x}/k_{0}$ and $k_{y}/k_{0}$ for two
sets of spin-orbit parameters $\alpha_1=0.4, \beta_1=0.4,
\beta_{3}=0.3$ (top panel) and $\alpha_1=0.2, \beta_1=0.8,
\beta_{3}=0.3$ (bottom panel). Note the saddle points correspond to
the most prominent van Hove singularities in the joint density of
states (and conductivity) in Fig.~4. The van Hove singularities are
at about $1.4E_0$ ($k_x=k_y$ in the momentum space) in the top frame
of Fig.~5 and at about $2E_0$ ($k_x=k_y$) and $0.9E_0$ ($k_x=-k_y$)
in the bottom.

\section{Discussion}

The optical conductivity is often characterized by the joint density
of states, $D(\omega )$, which has a finite onset at small energies.
This is well known in the graphene literature where interband
transitions start exactly at a photon energy equal to twice the
chemical potential. Here this still holds approximately in all the
cases considered in Fig.~4 except for the solid red curve in the two
left side frames. In this case $\alpha_1 =\beta _{1}=0.4$ and
$\beta_{3}$ is non zero. If $\beta _{3}$ is small the energy
$\sqrt{S_{1}^{2}+S_{2}^{2}}$ would be approximately equal to
$\sqrt{2}k\alpha_{1}\sqrt{1+\sin 2\theta }$, which is zero for
$\theta =3\pi/4$, the critical angle in the spin texture of the
lower left frame of Fig.~1 for which all spins are locked in this
direction. This means that only the quadratic term ${\hbar
^{2}k^{2}}/{(2m)}$ and cubic Dresselhaus term contribute to the
dispersion curve in this direction and there is no linear (in $k$)
graphene-like contribution. Thus, the onset of the interband optical
transition no longer corresponds to $\omega =2\mu $.

Considering the case of positive $\mu$, for the direction
$\theta=3\pi/4$, $(k/k_{0})^{2}/2+\beta _{3}(k/k_{0})^{3}$ is the
dominant contribution to the energy which is equal to $\mu/E_{0}$
and the minimum photon energy is now $2\beta _{3}(k/k_{0})^{3}$,
which could be very small as is clear from the figure. For negative
values of $\mu $ the onset is closer to $2|\mu|/E_{0}$ because in
this case the momentum at which the chemical potential crosses the
band dispersion is given by
$(k/k_{0})^{2}/2-\alpha_1(k/k_{0})=-\mu/E_{0}$ (the cubic term is
ignored because it is subdominant for small $k/k_{0}$ compared to
the linear term). Now the photon energy onset will fall above
$2|\mu|/E_{0}$, at a value dependent on $\alpha_1$.

While the optical conductivity Eq.~(\ref{Cond}) requires a non-zero
joint density of states Eq.~(\ref{DOS}), the additional weighting of
$(V_{x}S_{2}+V_{y}S_{1})^{2}$ in $\sigma _{xx}(\omega )$ can
introduce considerable changes to its $\omega$ dependence \cite{May}
as we see in Fig.~3 and Fig.~4. In the top frame of Fig.~3, $\beta
_{3}=0$ and there are no van Hove singularities because the Dirac
contribution to the dispersion curves simply increases with
increasing $k$. The solid black and dashed red curves both reduce to
the pure graphene case with onset exactly at $2\mu $ and flat
background beyond. The dotted red curve for mixed linear Dresselhaus
and Rashba is only slightly different. The onset is near but below
$2\mu $ and the background has increased in amplitude. It is also no
longer completely flat to high frequency; instead it has a kink near
$\hbar \omega /E_{0}\approx 1.7$ after which it drops. The
dash-dotted black curve for $\alpha_1 =0.4$ and $\beta _{1}=0.6$ has
changed completely with background reduced to near zero but with a
large peak corresponding to an onset which has shifted to a value
much less than $2\mu$. Finally for $\alpha_1 =\beta _{1}$ the entire
interband transition region is completely depleted as we know from
Eq.~(\ref{Cond}).

In Fig.~4 there is (non-zero) cubic Dresselhaus coupling present.
The solid red curves, for which $\alpha_1 =\beta_{1}$ but with
$\beta _{3}=0.3$ illustrate that the conductivity on the left
(positive $\mu$) is non-zero, and $\beta_3 = 0$ is necessary for a
vanishing interband conductivity at all photon energies. We see,
however, that these transitions have been greatly reduced below what
they would be in graphene for all photon energies except for a
narrow absorption peak at $\omega $ much less than $2\mu$. For
negative values of $\mu$, on the other hand, even with $\beta_3 \ne
0$ the conductivity is zero.

The experimental observation of such a narrow low energy peak
together with high energy van Hove singularities could be taken as a
measure of nonzero $\beta_{3}$. It is interesting to compare these
curves for the conductivity with the joint density of states (lower
frames). The color and line types are the same for both panels. The
onset energy as well as energies of the van Hove singularities are
unchanged in going from the joint density of states to the
conductivity. Also, as is particularly evident in the dotted black
and short dashed red curves the $1/\omega$ factor in $\sigma
_{xx}(\omega )$ leads to a nearly flat background for the
conductivity as compared with a region of nearly linear rise in the
density of states. This is true for both positive and negative
values of $\mu$.

In conclusion we have calculated the interband longitudinal
conductivity as a function of photon energy for the case of combined
Rashba and Dresselhaus spin-orbit coupling. We have also considered
the possibility of a cubic Dresselhaus contribution. We find that in
the persistent spin helix state when the spins are locked at an
angle of $3\pi/4$ independent of momentum, which arises when the
linear Rashba coupling is equal to the linear Dresselhaus coupling,
the interband optical transitions vanish and there is no finite
energy absorption from these processes. Only the Drude intraband
transitions will remain. When the cubic Dresselhaus term is nonzero
the cancelation is no longer exact but we expect interband
absorption to remain strongly depressed for photon energies above
$2\mu$ as compared, for example, to the universal background value
found in single layer graphene. We propose interband optics as a
sensitive probe of the relative size of Rashba and Dresselhaus spin
orbit coupling as well as cubic corrections.

\section{Methods}

The optical conductivity is given by
\begin{eqnarray}
\sigma _{xx}(\omega )=\frac{e^{2}}{i\omega }\frac{1}{4\pi ^{2}}
\int_{0}^{k_{cut}}kdkd\theta T\sum_{l}Tr\langle
\hat{v}_{x}\widehat{G}(\mathbf{k,}\omega
_{l})\hat{v}_{x}\widehat{G}(\mathbf{k,}\omega _{n}+\omega
_{l})\rangle _{i\omega _{n}\rightarrow \omega +i\delta }.
\end{eqnarray}
Here $T$ is the temperature and $Tr$ is a trace over the $2\times2$
matrix, and $\omega _{n}=(2n+1)\pi T$ and $\omega _{l}=2l\pi T$ are
the Fermion and Boson Matsubara frequencies respectively with $n$
and $l$ integers. To get the conductivity which is a real frequency
quantity, we needed to make an analytic continuation from imaginary
$i\omega _{n}$ to $\omega + i\delta$, where $\omega$ is real and
$\delta $ is an infinitesimal. The velocity operators $\hat{v}_{x}$
and $\hat{v}_{y}$ are given by
\begin{eqnarray}
\hat{v}_{x} &=&\frac{\partial H_{0}}{\hbar \partial k_{x}}=V_{I}\hat{I}+V_{x}\hat{\sigma}
_{x}+V_{y}\hat{\sigma} _{y}  \notag \\
\hat{v}_{y} &=&\frac{\partial H_{0}}{\hbar \partial k_{y}}=V_{I}^{\prime
}\hat{I}+V_{x}^{\prime }\hat{\sigma} _{x}+V_{y}^{\prime }\hat{\sigma} _{y}.
\end{eqnarray}
Here $V_{I} = \hbar k_{x}/{m}$, $V_{x}=(\beta _{1} - \beta _{3} k_{y}^{2})/\hbar$,
$V_{y}= (-\alpha_1 + 2\beta _{3}k_{y}k_{x})/\hbar$,
$V_{I}^{\prime} = \hbar k_{y}/{m}$, $V_{x}^{\prime }=(\alpha_1 - 2\beta _{3} k_{y} k_{x})/\hbar $ and
$V_{y}^{\prime} = (-\beta _{1} + \beta_3 k_{x}^{2})/\hbar$.

The Green's function can be written as \cite{grimaldi06}
\begin{equation}
\widehat{G}(\mathbf{k},\omega _{n})=\frac{1}{2}\sum_{s=\pm
}(\hat{I}+s\mathbf{F}_{\mathbf{k}}\cdot \hat{\mathbf{\sigma
}})G_{0}(\mathbf{k},s,\omega_{n})
\end{equation}
where $\mathbf{F}_{\mathbf{k}} =
(S_{1},-S_{2},0)/\sqrt{S_{1}^{2}+S_{2}^{2}}$,
\begin{equation}
G_{0}(\mathbf{k},s,\omega _{n})=\frac{1}{i\hbar \omega _{n}+\mu -\frac{\hbar
^{2}k^{2}}{2m}-s\sqrt{S_{1}^{2}+S_{2}^{2}}}
\end{equation}
and
\begin{eqnarray}
S_{1} &=&(\alpha_1 k_{y}+\beta _{1}k_{x}-\beta _{3}k_{x}k_{y}^{2})  \notag \\
S_{2} &=&(\alpha_1 k_{x}+\beta _{1}k_{y}-\beta _{3}k_{y}k_{x}^{2})
\end{eqnarray}
The wave function is given by
\begin{equation}
\Psi _{\mathbf{k,}\pm } | 0> =\frac{1}{\sqrt{2}} \biggl[ c_{{\bf k},
\uparrow}^\dagger | 0> \pm
\frac{S_{1}-iS_{2}}{\sqrt{S_{1}^{2}+S_{2}^{2}}} c_{{\bf k},
\downarrow}^\dagger | 0> \biggr], \label{eigenstates_old}
\end{equation}
with corresponding eigenvalues
\begin{equation}
E_{\pm }=\frac{\hbar ^{2}k^{2}}{2m}\pm \sqrt{S_{1}^{2}+S_{2}^{2}}.
\label{eigenvalues}
\end{equation}
Here $c_{{\bf k}, \uparrow}^\dagger(c_{{\bf k},
\downarrow}^\dagger)$ creates a particle with momentum $\bf k$ and
spin up (down). The spin expectation values work out to be
\begin{eqnarray}
S_{x}&=&\frac{\hbar }{2}\langle \Psi _{\mathbf{k,}\pm }|\sigma
_{x}|\Psi _{\mathbf{k,}\pm }\rangle =\pm \frac{\hbar}{2}\frac{S_{1}}{\sqrt{S_{1}^{2}+S_{2}^{2}}}  \notag \\
S_{y}&=&\frac{\hbar }{2}\langle \Psi _{\mathbf{k,}\pm }|\sigma
_{y}|\Psi _{\mathbf{k,}\pm }\rangle =\pm \frac{\hbar}{2}\frac{-S_{2}}{\sqrt{S_{1}^{2}+S_{2}^{2}}}  \notag \\
S_{z}&=&\frac{\hbar }{2}\langle \Psi _{\mathbf{k,}\pm }|\sigma
_{z}|\Psi _{\mathbf{k,}\pm }\rangle = 0.
\end{eqnarray}

These formulas allow us to calculate the spin texture, as well as the optical conductivity as given
in Eq. (\ref{Cond}).

\section{Acknowledgments}

\begin{acknowledgments}
This work was supported by the Natural Sciences and Engineering
Research Council of Canada (NSERC), the Canadian Institute for
Advanced Research (CIFAR), and Alberta Innovates.
\end{acknowledgments}

\section{Author Contributions}

ZL carried out the calculations, and all authors, ZL, FM, and JPC
contributed equally to the development of the work.

\section{Competing financial interests}

The authors declare no competing financial interests.

\end{document}